\documentclass[pra,aps,draft]{revtex4}

\newcommand{\ket}[1]{|#1\rangle}

\newcommand{\scpr}[2]{\langle#1|#2\rangle}

\begin{document} 


\title{The Group-Theoretical Classification of Some Multiparticle
  States in the Presence of Magnetic Field and Periodic Potential}

\author{Wojciech Florek\thanks{electronic mail: florek@amu.edu.pl}}

\affiliation{A. Mickiewicz University, Institute of Physics\\
  ul. Umultowska 85, 61--614 Pozna\'n, Poland}

\begin{abstract}
 The group-theoretical classification of multiparticle states 
 (pairs of particles and charged excitons $X^\pm$) is based on considerations
 of products of irreducible projective representations of the 
 two-dimensional  translation group. The states of a pair particle-antiparticle
 are non-degenerate, whereas, for a given Born--von K\'arm\'an 
 period $N$, degeneracy of pair states is $N$ and
 three-particle states are $N^2$-fold degenerated. The symmetrization of states with 
 respect to particles transposition is considered.  
 Three symmetry adapted bases for trions are considered: 
 (i) the first is obtained from a direct conjugation of three representations;
 (ii) in the second approach the states of a electrically neutral pair 
 particle-antiparticle are determined in the first step; (iii) the third 
 possibility is to consider a pair of identical particles in the first step. 
 In the discussion presented the Landau gauge ${\mathbf A}=[0,Hx,0]$ is used, 
 but it is shown that the results 
 obtained are gauge-independent. In addition the relation between changes of
 a chosen gauge and {\em local}\/ basis transformations are discussed.
\end{abstract}

\maketitle
\section{Introduction}

The quantum Hall effect and high temperature superconductivity have given
raise to interest in properties of the two-dimensional electron gas 
subjected to electric and
magnetic fields. The observation of (negatively) charged 
excitons \cite{shields} has recalled a forty-year old concept of excitons ``trions''
or ``charged excitons'' introduced by Lampert in 1958 \cite{lamp}.
Recently, such excitons, consisting of two holes and an electron or two 
electrons and hole (denoted $X^\pm$, respectively), have been investigated 
both experimentally and theoretically \cite{exper,theor,dzy}.

In this paper classification based on translational symmetry in the presence 
of a periodic potential and an external magnetic field is presented. To 
perform this task the so-called magnetic translation operators, introduced by 
Brown \cite{brown} and Zak \cite{zak}, are used. These operators commute with 
the standard Hamiltonian of an electron in the magnetic field 
${\mathbf H}=\nabla\times{\mathbf A}$ and a periodic potential 
$V({\mathbf r})$ 
\begin{equation}\label{ham}
  {\mathcal H} =
  \frac{1}{2m}\left({\mathbf p}+\frac{e}{c}{\mathbf A}\right)^2
     + V({\mathbf r})\,.
\end{equation}
This paper exploits the fact that after imposing the Born--von K\'arm\'an (BvK)
periodic conditions the magnetic translations form a finite-dimensional  projective 
representation of the 2D translation group. Kronecker products of irreducible 
projective representations can be applied to description of multiparticle 
states \cite{flo99}.

The aim of this work is to present classification of three-particle
states (strictly speaking states of particle-particle-antiparticle systems)
when at the first step states of a pair of identical (charged) particles
are constructed. Such an approach allow to discuss pair states, what may 
be important in considerations of high-$T_{\text{c}}$ superconductors,
where Cooper pairs are confined to Cu-O planes. A pair particle-antiparticle
is also considered and a particle is added in the 
next step, what 
leads to the particle-particle-antiparticle system. States of such systems
are also determined by means of direct conjugation of three representations. 
More detailed discussion of trions is also presented in another
author's article \cite{jmpnew}.

Investigating problems, which involve the magnetic field ${\mathbf H}$ 
determined
by the vector potential ${\mathbf A}$, one has to keep in mind that some
results may depend on a chosen gauge, though physical properties should
be gauge-independent. Two gauges are most frequently used in description of 
the 2D electron systems: the Landau gauge with ${\mathbf A}=[0,Hx,0]$ and the
antisymmetric one with ${\mathbf A}=({\mathbf H}\times{\mathbf r})/2$. The 
relations between these gauges were discussed in the earlier 
article \cite{jmpold}. For the sake of simplicity the considerations are limited 
the Landau gauge, but the previous results \cite{jmpold} and the concept of
rays \cite{weyl} together with application
of a {\em local}\/ basis transformation enable us to show that 
classification obtained here is gauge-independent.  
The presented results correspond
to the limit of high magnetic fields, \textit{i.e.}\ there is no Landau level
mixing. 

\section{Magnetic translations as projective representations}

Magnetic translation operators $T({\mathbf R})$
commuting with the Hamiltonian (\ref{ham}) 
form a projective representation of the translation 
group ${\mathcal T}$ \cite{brown}, \textit{i.e.}
 \[
  T({\mathbf R}_1)T({\mathbf R}_2)=T({\mathbf R}_1+{\mathbf R}_2)
  \mu({\mathbf R}_1,{\mathbf R}_2)\,;
 \]
 ${\mathbf R}_1,{\mathbf R}_2\in {\mathcal T}$,
 $\mu({\mathbf R}_1,{\mathbf R}_2)\in\text{U}(1)$.
 It should be stressed that due to the presence of a magnetic field ${\mathbf H}$
all interesting physical (as well as mathematical) features take place in 
the plane perpendicular to ${\mathbf H}$. Therefore, we assume
${\mathbf H}=H\hat{{\mathbf z}}$ and a two-dimensional crystal
lattice determined by vectors ${\mathbf a}_1$, ${\mathbf a}_2$ lying in the
$xy$-plane is considered.
The concept of magnetic translations can be generalized to a local gauge of 
the vector potential ${\mathbf A}$, $d$-dimensional lattices, and the spatially 
inhomogeneous magnetic field \cite{jmpold,flomc,thom}. The periodic 
boundary conditions give rise to the flux quantization:  
 \[
  q\phi=\frac{l}{n}\,,\qquad n|N\,,\; \gcd(l,n)=1\,;
 \]
  $q=Q/e$, where $Q$ is a particle charge, $\phi$ is a flux through the unit 
lattice cell, and $N$ is the Born--von~K\'arm\'an period. All possible pairs 
$(l,n)$ label different irreducible representations of the translation group 
${\mathcal T}$. Two notes are in a place. At first, please note that the 
formula presented means, it fact, quantization of the product $q\phi$ and 
should be applied to systems containing particles with $|q|>1$ \cite{jmpold}.
In their original works Brown\cite{brown} and Zak\cite{zak} dealt with 
electrons only. In the case $n=N$ each $l$, mutually prime with $N$, 
determines
a projective irreducible representation (irrep) in the unique way. 
In the other cases ($1\le n<N$) there
is an additional pair of indices ${\mathbf k}\equiv[k_1,k_2]$ 
($0\le k_j<(N/n)$, $j=1,2$), so there are
$(N/n)^2$ $n$-dimensional nonequivalent projective irreps of ${\mathcal T}$.
Each of these irreps has the same factor system 
$\mu({\mathbf R}_1,{\mathbf R}_2)$ determined by the pair $(l,n)$ \cite{BB}.
This property is related to the concept of the so-called magnetic cells
and magnetically periodic conditions \cite{brown,flomc}.

\section{Product of projective irreps}

Assuming the Landau gauge, ${\mathbf A}=[0,Hx,0]$, the finite-dimensional
projective irrep of the 2D translation group can be chosen as ($n=N$)
 \[
  D^l_{jk}[n_1,n_2]=\delta_{j,k-n_2}\omega_N^{ln_1j}\,,
 \]
 where $\omega_N=\exp(2\pi\text{i}/N)$ and all expressions are calculated 
modulo $N$; vectors of the translation group ${\mathcal T}\ni{\mathbf R}
=n_1{\mathbf a}_1+n_2{\mathbf a}_2$ are replaced by their coordinates, 
$n_1$ and $n_2$, in the crystal basis.  In the terms of the basis vectors
of the irrep we obtain
 \[
   D^l[n_1,n_2]\ket{w}=\omega_N^{ln_1(w-n_2)} \ket{w-n_2}\,.
  \]
 The other special case, $n=1$, leads to the $N^2$ standard (vector) irreps 
of the translation group labeled by wave vector ${\mathbf k}$, \textit{i.e.}
 \begin{equation}\label{Dk} 
   D^{\mathbf k}[n_1,n_2]=\omega_N^{k_1n_1+k_2n_2}\,, 
 \end{equation}
 where ${\mathbf k}=[k_1,k_2]$, $k_1,k_2=0,1,2,\dots,N-1$.

The Kronecker product of representations $D^l$ (for any $n$) corresponds to 
the addition of charges (for a fixed $H$) \cite{flo99,flopro}.
In particular
 \begin{eqnarray}
   D^l\otimes D^{-l} &=& \bigoplus_{\mathbf k} D^{\mathbf k}\,;\label{pape}\\
   D^l\otimes D^l\otimes D^{-l}&=&N^2 D^l\quad \text{ for } n=N\,.\label{teq}
 \end{eqnarray}
 In the further considerations we limit ourselves to the case, when one-particle
states are described by representations with $n=N$. It means that crystal and
magnetic periods \cite{jmpold} are identical and we consider products of 
representations $D^l$ for $l$ mutually prime with $N$. 

\section{States of pairs}

\subsection{States of particle-antiparticle pair\label{pap}}

The particle-antiparticle pair is very simple to consider. Taking into
account Eq.~(\ref{pape}) one can construct symmetry adapted basis
consisting of vectors  
 \[
  \ket{0}_{\mathbf k}=N^{-1/2}\sum_{s=0}^{N-1} 
    \omega_N^{sk_2} \ket{s}_+\ket{s-xk_1}_-\,,
 \] 
 where $xl=1\bmod N$ ($x=l^{-1}\bmod N$). The vector 
$\ket{0}_{\mathbf k}$ transform as the basis vector of the one-dimensional
(vector) representation $D^{\mathbf k}$ determined in Eq.~(\ref{Dk}). 
The subscript `+' (`--') denotes states of a particle (an antiparticle,
respectively).

\subsection{Pair of identical particles\label{idenp}}

In the case of a pair of identical particles, \textit{i.e.}\ considering
the product $D^l\otimes D^l$, one has to distinguish cases of odd and
even $N$. The first case ($N$ odd) is easy to solve since 
 \[
    D^l\otimes D^l = N D^{2l}\,.
 \]
 One of possible forms of the irreducible basis is
 \begin{equation}\label{podd}
  \ket{s}_2^r = \ket{s+r}_+ \ket{s-r}_+\,,\qquad s,r=0,1,2,\dots,N-1\,,
 \end{equation}
 where the subscript `2' indicates states of a pair and $r$ is the repetition 
index.  

 The case $N=2M$ leads to the representations product in the
following form 
 \[
    D^l\otimes D^l 
 = M (D^{l;[0,0]}\oplus D^{l;[1,0]}\oplus D^{l;[0,1]}\oplus D^{l;[1,1]})\,.
 \]
 The representations $D^{l;[k_1,k_2]}$, $k_1,k_2=0,1$ form
a complete set of nonequivalent $M$-dimen\-sional projective irreps of 
${\mathcal T}$. They can be expressed by a general formula \cite{BB}
 \[
  D^{l;[k_1,k_2]}[n_1,n_2]\ket{s}
  =\omega_N^{ln_1(2s-2\eta+k_1)}(-1)^{\xi k_2}\ket{s-\eta}_+\,,
 \] 
 where $k_1,k_2=0,1$, $\eta=n_2\bmod M$, $\xi=1$ for 
$M\leq s-n_2\leq 2M$ and $\xi=0$ otherwise. 
The irreducible bases can be chosen in the following way 
 \begin{equation} \label{baskk}
    \ket{s}_2^{r;[k_1,k_2]}=2^{-1/2} 
     \left(\ket{s+r}_+\ket{s-r+k_1}_+
    +(-1)^{k_2}\ket{M+s+r}_+\ket{M+s-r+k_1}_+\right)
 \end{equation} 
   with $s,r = 0,1,2,\dots,M-1$, where $s$ labels vectors and $r$
is the repetition index.  

\subsection{Symmetrization of states}

Having determined the basis $\ket{s}_2$ for pairs of identical particle it 
is natural and obvious to investigate the symmetry properties related 
with the transposition of particles. As in all above problems the case of 
odd $N=2M+1$ is quite easy. It follows from Eq.~(\ref{podd}) that $r=0$ leads 
to symmetric states
 \[
   \ket0_+\ket0_+,\, \ket1_+\ket1_+,\,\dots, \ket{N-1}_+\ket{N-1}_+\,.
 \]
 The other $N^2-N=N(N-1)$ vectors correspond to $N-1=2M$ representations 
labeled
by $r=1,\dots,N-1$. To construct symmetric (antisymmetric) states it is 
enough to take combinations
  \[
     \ket{s}_2^{r\pm} = 2^{-1/2}(\ket{s+r}_+\ket{s-r}_+\pm \ket{s-r}_+\ket{s+r}_+)\,,
  \]
  where $r=1,2,\dots,M$ now.

In the case of even $N=2M$ the product $D^l\otimes D^l$ decomposes into
$M$ copies of four $M$-dimensional projective representations. Since
the symmetrization of states has a slightly different form in each
of these cases then they are considered separately.

The non-symmetrized basis of the representation $D^{l;[0,0]}$ is given
by Eq.~(\ref{baskk}) as
 \[
  \ket{s}_2^{r;[0,0]} = 2^{-1/2}\left(\ket{s+r}_+\ket{s-r}_+
      +\ket{M+s+r}_+\ket{M+s-r}_+\right)\,.
 \] 
 The case $r=0$ again gives $M$ symmetric states. To consider the other 
$M-1$ representations one has to check the parity of $M$. If $M=2\mu$,
then for $r=\mu$ we have $M-r=r$ and $M+r=N-r$, so in this case all
vectors are symmetric. The other vectors form symmetric and antisymmetric
states in the standard way and it also concerns the case of odd $M$. 

The `additional' symmetric state found above for $N=4\mu$ is `lost'
considering $D^{l;[1,0]}$. Its irreducible basis given by Eq.~(\ref{baskk})
 \[
  \ket{s}_2^{r;[1,0]} = 2^{-1/2}\left(\ket{s+r}_+\ket{s-r+1}_+
      +\ket{M+s+r}_+\ket{M+s-r+1}_+\right)
 \]
 have no symmetric states for $M=2\mu$, but for odd $M=2\mu-1$ and $r=\mu$
the above formula reads
 \[
  \ket{s}_2^{\mu;[1,0]} = 2^{-1/2}\left(\ket{s+\mu}_+\ket{s-\mu+1}_+
      +\ket{s-\mu+1}_+\ket{s+\mu}_+\right)\,,
 \]
 so these states are symmetric.

The similar considerations have to be performed for representations 
$D^{l;[k_1,1]}$ with special attention to the fact that 
 \begin{eqnarray*}
  \ket{s-r}_+\ket{s+r}_+&-&\ket{M+s-r}_+\ket{M+s+r}_+ \\
  &&~~~~=-\left(\ket{s+r'}_+\ket{s-r'}_+
      +\ket{M+s+r'}_+\ket{M+s-r'}_+\right)\,,
 \end{eqnarray*}
where $r=0,1,\dots,M-1$ and $r'=M-r$.

\section{States of trions}

Trions $X^\pm$ were introduced in 1958 by Lampert \cite{lamp} as excitons
consisting of two holes and an electron ($X^+$) or two electrons and a
hole ($X^-$). This concept can be generalized to any system of
two particles and one antiparticle. 

\subsection{Direct conjugation}

In this case the Kronecker product of three 
representations presented in Eq.~(\ref{teq}) 
should be considered. It can be verified that the symmetry adapted vectors 
can be constructed as 
 \begin{equation}\label{trion}
   \ket{w}_t^{pq}=\ket{w+p}_+\ket{w+q}_+\ket{w+p+q}_-\,,
 \end{equation}
 where a pair $(p,q)$, $p,q=0,1,2,\dots,N-1$, plays a role of a repetition
index and the subscript $\pm$ denotes vectors of the irreps $D^{\pm l}$,
respectively.

  For $p=q$ the states obtained are symmetric with respect to the 
transposition of identical particles. In the other
cases ($p\neq q$) symmetric and antisymmetric combinations can be formed:
 \[
   \ket{w}_t^{pq\pm}=2^{-1/2}(\ket{w}_t^{pq}\pm\ket{w}_t^{qp})\,,
 \]
 where now $q>p=0,1,\dots,N-1$. One obtains $N(N-1)/2$ 
antisymmetric states $\ket{w}_t^{pq-}$ and $N(N+1)/2$ symmetric ones
$\ket{w}_t^{pq+}$. 

\subsection{Conjugation via the neutral pair}

In this case we start from the particle-antiparticle pair
considered in Sec.~\ref{pap} and the product of three representations
is written as
 \[
  \left(D^l\otimes D^{-l}\right) \otimes D^l = 
  \bigoplus_{\mathbf k} D^{\mathbf k} \otimes D^l\,.
 \]
 Since for each ${\mathbf k}$ one has $D^{\mathbf k}\otimes D^l=D^l$,
 then each such product yields states 
 \[ 
  \ket{w}_t^{\mathbf k}= 
    \omega_N^{-wk_2} \ket{0}_{\mathbf k}\ket{w-xk_1}_+\,.
  \]
 Equations obtained lead to the final expression
($xl = 1 \bmod N$, $N^{-1/2}$ is a normalization factor)
 \[
  \ket{w}_t^{\mathbf k}= N^{-1/2}\omega_N^{-wk_2}
 \sum_{s=0}^{N-1} \omega_N^{sk_2} \ket{s}_+\ket{s-xk_1}_-\ket{w-xk_1}_+\,.
 \]
 In such a state there is a kind of symmetry between a particle and 
antiparticle, but there is no symmetry between
two identical particles in the trion $X^+$. Since there are $N^2$ trion 
states labeled by $w$ then it is possible to construct states symmetric and 
antisymmetric with respect to transposition of particles. 

Scalar products with the vectors obtained previously are 
 \[
   {}_{\qquad t}^{[k_1,k_2]}\scpr{w} {w}_t^{pq} 
   = N^{-1/2}  \omega_N^{-pk_2} \delta_{q+xk_1,0}\,.
 \]
 In the simplest case $N=2$ this formula yields (the unique representation is
obtained for $l=x=1$); 
 \begin{eqnarray*}
  \ket{w}_t^{00} = 2^{-1/2}\left(
    \ket{w}_t^{[0,0]} + \ket{w}_t^{[1,0]} \right) \,,&\qquad&
  \ket{w}_t^{01} = 2^{-1/2}\left(
    \ket{w}_t^{[0,0]} - \ket{w}_t^{[1,0]} \right)\,, \\
  \ket{w}_t^{11} = 2^{-1/2}\left(
    \ket{w}_t^{[0,1]} - \ket{w}_t^{[1,1]} \right)\,, &&
  \ket{w}_t^{10} = 2^{-1/2}\left(
    \ket{w}_t^{[0,1]} + \ket{w}_t^{[1,1]} \right) \,.
  \end{eqnarray*}
 Symmetrization of the formulas in the right column leads to the following 
expressions 
  \begin{eqnarray*}
   \ket{w}_t^{[0,1]+} &=& \left( \ket{w}_t^{[0,0]} - \ket{w}_t^{[1,0]} 
    + \ket{w}_t^{[0,1]} + \ket{w}_t^{[1,1]} \right)/2\,, \\
   \ket{w}_t^{[0,1]-} &=& \left( \ket{w}_t^{[0,0]} - \ket{w}_t^{[1,0]}
    - \ket{w}_t^{[0,1]} - \ket{w}_t^{[1,1]} \right)/2\,. 
  \end{eqnarray*}

\subsection{Conjugation via the pair of identical particles}

In this case we use the results of Sec.~\ref{idenp}, so we start from 
products 
  \[
   D^{2l}\otimes D^{-l}= N D^l\,, \qquad 
   D^{l;[k_1,k_2]}\otimes D^{-l} = M D^l\,. 
  \]
Note that in both cases we obtain copies of the representation $D^l$.

 The first case corresponds to $N$ odd and again is very simple. 
One of possible choice of basis vectors is 
  \[
   \ket{w}_t^v=\ket{w+v}_2 \ket{w+2v}_-\,,\qquad
   w=0,1,2,\dots,N-1\,,
  \]
  where $v=0,1,2,\dots,N-1$ is the repetition index. 
Taking into account the previous results 
 \[
  \ket{w}_t^{r,v}=\ket{w+v}_2^r\ket{w+2v}_- 
     = \ket{w+v+r}_+ \ket{w+v-r}_+ \ket{w+2v}_-\,.
 \] 
  The states
 \[
    \ket{w}^{pq}_t = \ket{w+p}_+\, \ket{w+q}_+\, \ket{w+p+q}_-
  \]
 obtained previously correspond
to $p=v+r$ and $q=v-r$ (calculated mod~$N$). Note that for $N=2M$ 
this relations can not be inverted since $v=(p+q)/2$ and $r=(p-q)/2$ have
no solutions mod~$N$ for odd $p+q$ and $p-q$, respectively.

In the case $N=2M$ considerations are a bit more difficult, but 
since one knows the final results, states $\ket{w}_t^{pq}$, then they
may serve as a useful hint. There are $N^2=4M^2$ different bases labeled by
$p,q=0,1,2,\dots,N-1$, however if $p'=p+M$ and $q'=q+M$ then states
$\ket{w}_t^{pq}$ and $\ket{w}_t^{p'q'}$ have the same third element in the
tensor products because 
 \[
    \ket{w}^{p'q'}_t =  
  \ket{w+p+M}_+ \ket{w+q+M}_+ \ket{w+p+q}_-\,.
 \]
 Therefore, they can be gathered into $2M^2=NM$ pairs
 \[
  \ket{w}_t^{pq}\qquad \mbox{and}\qquad \ket{w}_t^{(p+M)(q+M)}\,. 
 \]
 The ranges of indices have to be chosen in such a way that pairs $pq$
and $(p+M)(q+M)$ run over two separate sets. For each pair $p,q$ one 
can form two new bases, labeled by `+' and `--', respectively,
 \begin{equation}\label{newbas}
  \ket{w}_t^{pq\pm}= \ket{w}_t^{pq}\pm \ket{w}_t^{(p+M)(q+M)}\,, 
 \end{equation}
 so
 \[
    \ket{w}_t^{pq\pm} = \ket{w+p+q}_- 
    \left(\ket{w+p}_+ \ket{w+q}_+
    \pm \ket{w+p+M}_+\ket{w+q+M}_+\right) \,.
 \]
 These vectors have the second part (in parentheses) in the form resembling
Eq.~(\ref{baskk}). Hence, one has to relate the repetition index $r$ 
and a label $u$ of a vector $\ket{u}_-$ of the 
representation $D^{-l}$ with the repetition indices $p,q$ and $\pm$ in
Eq.~(\ref{newbas}). The solution of this problem leads to a proper choice of 
pairs $pq$. This has been discussed
in more details elsewhere \cite{flo2001}.

\section{Other gauges}

A trivial factor system $\theta$ is determined by any mapping 
$\phi\colon G\to\text{U}(1)$ as 
 \[
  \theta({\mathbf R},{\mathbf R}')=\phi({\mathbf R})\phi({\mathbf R})\,/\,
  \phi({\mathbf R}+{\mathbf R}')\,.
 \]
 Let $D$ be a projective representation with a factor system $\mu$.
A new representation $D'$ determined as $D'({\mathbf R})
=\phi({\mathbf R})D({\mathbf R})$ has an equivalent factor system 
$\mu'=\theta\mu$. On the other hand, for any unitary
operator $S$ and $D''({\mathbf R})=SD({\mathbf R})S^{-1}$ one obtains 
 \[
   D''({\mathbf R})D''({\mathbf R}')=\mu({\mathbf R},{\mathbf R}')
    D'' ({\mathbf R}+{\mathbf R}')\,.
 \]
 Therefore, \textit{equivalent}\/ representations have \textit{identical}\/ factor
systems \cite{alt}. It means that equivalent factor systems, corresponding 
to different gauges \cite{jmpold}, lead to \textit{nonequivalent}\/ representations.
However, it will be shown below, that these factors (so gauges, too) are related
to a \textit{local}\/ transformation of state space.

The most popular gauges, \textit{i.e.}\ the Landau and
the symmetric ones, are the special cases of the so-called linear gauge \cite{thom},
which has the form 
 \[
  {\mathbf A}({\mathbf r})=H[-\beta y, (1-\beta) x]\,.
 \]
The relation between the vector potential ${\mathbf A}$ and the factor system
of a projective representation \cite{jmpold} yields that $\beta=0$
corresponds to the representation considered above, whereas $\beta=1$ and 
$\beta=1/2$, determining the other form of the Landau gauge and the 
(anti)symmetric one ${\mathbf A}=({\mathbf H}\times{\mathbf r})/2$ \cite{thom,flo97}.
Thus relation leads to matrices
 \begin{eqnarray*}
   {^1}\!D^l_{jk}[n_1,n_2] &=& \delta_{j,k-n_2}\omega_N^{ln_1k}\,,\\
   {^{1/2}}\!D^l_{jk}[n_1,n_2] &=& \delta_{j,k-n_2}\omega_N^{ln_1(j+k)/2}\,.
 \end{eqnarray*}
  In a general case one obtains   
  \[
   {^\beta}\!D^l_{jk}[n_1,n_2]=
   \delta_{k-j,n_2} \omega_N^{ln_1[(1-\beta) j+\beta k]}
   = \delta_{k-j,n_2} \omega_N^{ln_1(j+\beta n_2)}
   = \omega_N^{ln_1n_2\beta}\,{^0}\!D^l_{jk}[n_1,n_2]\,.
  \]
  The action of operators ${^\beta}\!D^l$ for vectors $[n_1,0]$ and 
  $[0,n_2]$ on the basis vectors yields
 \[
  {^\beta}\!D^l[n_1,0]|j\rangle=\omega_N^{ln_1j}|j\rangle\,,\qquad
  {^\beta}\!D^l[0,n_2]|j\rangle=|j-n_2\rangle\,.
 \] 
Therefore, they behave in the same way for all real numbers
$\beta$. The differences can be only noticed for general translations
$[n_1,n_2]$ and, moreover, the non-zero matrix elements 
obtained for different 
$\beta$ appear in the same places (it is controlled by the 
$\delta_{k-j,n_2}$),
but are multiplied by different powers of $\omega_N^l$. Let us recall 
that projective representations are sometimes, especially in physics \cite{brown},
called {\em ray}\/
representations.  Since only a module of a bracket
$\langle j|k\rangle$ has a physical meaning, then vectors (complex
functions) $|j\rangle$ and $|k\rangle$ are determined up to factors
$\lambda\in {\mathrm U}(1)$ \cite{weyl}. Therefore, a chosen state
$|j\rangle$ represents in fact a ray, \textit{i.e.}\ the set 
$||j\rangle\rangle=\{\lambda|j\rangle\mid \lambda\in {\mathrm U}(1)\}$. 
In a general case
 \[
  {^\beta}\!D^l_{jk}[n_1,n_2] =\langle j|{^\beta}\!D^l[n_1,n_2]|k\rangle 
   =\delta_{k-j,n_2} \omega_N^{-ln_1j\beta} \omega_N^{ln_1j} \omega_N^{ln_1k\beta}\,.
 \]
 Therefore, replacing each vector $\ket{j}$ by an element of the same
ray $\ket{j'}=\omega_N^{-ln_1j\beta}$ one obtains
 \[
  \langle j'|{^\beta}\!D^l[n_1,n_2]|k'\rangle 
  =\omega_N^{ln_1(j-k)\beta}\langle j|{^\beta}\!D^l[n_1,n_2]|k\rangle 
  =\delta_{k-j,n_2} \omega_N^{ln_1j} ={^0}\!D^l_{jk}[n_1,n_2] \,.
 \]
 However, this transformation is {\em local}, because it depends 
 on $n_1$ and $n_2$, since $k-j=n_2$ for non-zero matrix elements.
If can be verified that equations obtained above are invariant under this
transformation, so all relations derived here for the Landau gauge are 
valid for other linear gauges. 

\section{Final remarks}
In this work the Kronecker products of irreducible projective
representations of the two-dimensional translation group are 
applied to to the group-theoretical classification of multiparticle states
in systems subjected to the external magnetic field. Such states 
may be useful in the considerations of quantum wells, high-$T_{\text{c}}$
superconductors and the fractional quantum Hall effect. In this last case,
there is special interest in trions, which are a special case of considered
here the system consisting of two particles and one antiparticle.

Due to internal structure of trion the degeneracy is higher and there are
many possibilities to construct states $\ket{w}_t$; some of them 
have been discussed above. In these simplified considerations there are
no interactions between trions or mixing of Landau levels and the spin
or angular momentum numbers have not been considered. Taking into account 
spins will allow to construct
states completely antisymmetric with respect to the permutational symmetry.
Such problem has been discussed lately by Dzyubenko \textit{et al.} \cite{dzy}
for the case of free trions (\textit{i.e.}\/ without a periodic potential, so
there is no discrete translational symmetry).
A sum of indices in the RHS of Eq.~(\ref{trion}), taking into account signs
of charges, is $(2w+p+q)-(w+p+q)=w$, what is equal to the index in the LHS 
of this equation. This is the same result as presented in \cite{dzy}, 
where
the total angular momentum projection of a trion equals $(n_1-m_1)+(n_2-m_2)
-(n_3-m_3)$ with $(n_j-m_j)$ being the total angular momentum projection for
holes ($j=1,2$) and an electron ($j=3$); $n$ and $m$ are the Landau 
level and the oscillator quantum numbers, respectively. It is interesting
that Dzyubenko \textit{et al.}\/ obtained their results in the antisymmetric
gauge ${\mathbf A}=({\mathbf H}\times{\mathbf r})/2$, whereas in the presented
considerations the Landau gauge has been used. It confirms that the physical
properties are gauge-independent. On the other hand, the actual form of wave 
functions is not discussed here, but the relations between
representations and their product are taken into account only. These relations
are independent of the matrix representations and, similarly, the form of
resultant basis is independent of the function form: for a given 
Born--von K\'arm\'an period $N$ and any linear gauge irreducible projective 
representations are $N$-dimensional and their action on basis vectors are 
similar (up to a factor system) \cite{brown,zak,jmpold,flo98}. This is proven
by taking into account the notion of rays and ray representations.

\section*{Acknowledgments}

This work is partially supported by the Committee for Scientific Research
(KBN) within the project No.\ 8~T11F~027~16.

\end{document}